\documentclass[twocolumn,english,aps,pra,reprint,nofootinbib,superscriptaddress]{revtex4-2}
\usepackage[T1]{fontenc}
\usepackage[latin9]{inputenc}
\usepackage{amsmath}
\usepackage{amssymb}
\usepackage{graphicx}
\usepackage{rotating}
\usepackage{tabularx}
\usepackage{braket}
\usepackage{color}
\usepackage{float}
\usepackage{hyperref}
\usepackage{multirow}
\hypersetup{
    colorlinks,
    linkcolor={black},
    citecolor={black},
    urlcolor={blue}
}

\makeatletter
\@ifundefined{textcolor}{}
{%
 \definecolor{BLACK}{gray}{0}
 \definecolor{WHITE}{gray}{1}
 \definecolor{RED}{rgb}{1,0,0}
 \definecolor{GREEN}{rgb}{0,1,0}
 \definecolor{BLUE}{rgb}{0,0,1}
 \definecolor{CYAN}{cmyk}{1,0,0,0}
 \definecolor{MAGENTA}{cmyk}{0,1,0,0}
 \definecolor{YELLOW}{cmyk}{0,0,1,0}
 \definecolor{PURPLE}{rgb}{0.7,0,0.7}
 \definecolor{dgreen}{rgb}{0,0.6,0}
}
\usepackage{xspace}

\usepackage{pslatex}

\usepackage{babel}

\usepackage{soul}

\makeatother

\begin{document}
\renewcommand{\figurename}{FIG.}
\renewcommand{\tablename}{TABLE}
\renewcommand{\appendixname}{APPENDIX}
\renewcommand{\arraystretch}{1.5}
\title{Laser cooling of a fermionic molecule}

\author{Jinyu Dai}
\affiliation{Department of Physics, Columbia University, New York, New York 10027-5255, USA}

\author{Qi Sun}
\affiliation{Department of Physics, Columbia University, New York, New York 10027-5255, USA}

\author{Benjamin C. Riley}
\affiliation{Department of Physics, Columbia University, New York, New York 10027-5255, USA}

\author{Debayan Mitra}
\email{dm3710@columbia.edu}
\affiliation{Department of Physics, Columbia University, New York, New York 10027-5255, USA}

\author{Tanya Zelevinsky}
\affiliation{Department of Physics, Columbia University, New York, New York 10027-5255, USA}
\date{\today}

\begin{abstract}
\noindent 
Only bosonic molecular species have been directly laser cooled to date, primarily due to an abundance of bosonic isotopes in nature and to their simpler hyperfine structure. Fermionic molecules provide new opportunities for ultracold chemistry, quantum simulation, and precision measurements. Here we report direct laser cooling of a fermionic molecular isotopologue, calcium monodeuteride (CaD). With a nuclear spin $I=1$, only 5 hyperfine states need to be addressed for rotational closure in optical cycling. These hyperfine states are unresolved for typical experimental linewidths. We present a method for efficiently producing alkaline-earth metal hydrides and deuterides. We demonstrate rotational closure and show magnetically assisted Sisyphus cooling in one dimension for a beam of CaD molecules. Our results indicate that the experimental complexity for laser cooling CaD is similar to that of calcium monohydride (CaH).  Laser cooling of CaD is a promising first step for production of ultracold and trapped atomic deuterium.

\end{abstract}

\maketitle

\section{Introduction}

Recent achievements in quantum-state control of molecules have demonstrated their potential for ultracold chemistry \cite{Yu_2022_chem_review,Ospelkaus2010KRb}, quantum simulation \cite{Christakis_2023_SingleSite_Molecules,RN18}, quantum computation \cite{Holland_2023_entanglement,Bao_2023_entanglement}, and precision measurements \cite{ACMENature18_ACMEIIeEDM,Roussy_2023_eEDM,ZelevinskyMitraPRA22_MoleculesFundamentalPhysics}. While fermionic molecules have been assembled from ultracold atoms \cite{Li_NatPhys2021_KRb_evaporation,Schindewolf_2022_evaporation,Park_NatPhys2023_NaLi_Magnetic_Trap}, to date all directly laser-cooled molecules have been exclusively bosonic. This stems partly from the fact that most abundant isotopes of neutral atoms in nature that constitute these molecules are bosonic, with the notable exceptions of beryllium and nitrogen. Additionally, a guiding principle for laser cooling candidate molecules is the simplicity of the hyperfine structure \cite{Shuman_LaserCoolingDiatomic_2010}. Most laser-cooled molecules are of the type $MX$, where $M$ is an optical cycling center with nuclear spin $I_M = 0$ and $X$ is an electronegative ligand with $I_X = 1/2$. The resulting hyperfine structure consists of only four states that need to be optically addressed for rotational closure of the optical cycling process \cite{ZelevinskyMitraPRA22_MoleculesFundamentalPhysics}. $I_M \ne 0$ for a fermionic species can significantly complicate the laser-cooling scheme \cite{Kogel_2021_BaF_fermionic}.

Fermionic molecules possess many favorable properties for quantum science applications. Being less prone to collisional loss than their bosonic counterparts due to the $p$-wave barrier \cite{Bause_2023_sticky_collisions}, fermionic molecules are an important ingredient in ultracold chemistry experiments. The combination of Fermi-Dirac statistics and long-range interactions of polar molecules can enable the realization of topological superfluid phases \cite{Cooper_2009_topological_superfluid} and lattice-spin models \cite{Micheli_2006_spin_toolbox}. A fermionic cycling center $M$ can be employed for precision measurements of nuclear-spin-dependent parity violation (NSD-PV) \cite{Norrgard_2019_NSDPV,Hao_2020_NSDPV} and axionlike dark matter searches \cite{Safronova_NewPhysicsAtoms_2018}. This  has motivated recent studies with fermionic $^{137}$BaF \cite{Kogel_2021_BaF_fermionic} and $^{171,173}$YbOH \cite{Zeng_2023_YbOH_complexHyperfine}, with the demonstration of  rotational closure for the latter. In order to overcome the hyperfine complexity that plagues the molecules listed above, we turn our attention to calcium monodeuteride (CaD)$-$the easiest fermionic molecule to laser cool in the near term.

Cold CaH and CaD isotopologues have astrophysical significance as they have been observed in stellar and interstellar media \cite{Frum_rotational_spectra_1993, Bernath2012spectrum, Alavi2018einsteincoeff}. Studying their ultracold chemical reactions in the laboratory would improve our understanding of fundamental chemical processes \cite{Tscherbul_2020PRL_CaH_Li_reaction}. Moreover, CaD is a promising precursor for producing ultracold atomic deuterium, in analogy to the proposals for CaH dissociation \cite{Lane_UltrcoldHydrogen_2015,Sun_PRR2023_Predissociation}. High-precision measurements with hydrogen and deuterium allow testing quantum electrodynamics \cite{boshier1987precision}, determining fundamental constants such as the proton charge radius \cite{doi:10.1126/science.abc7776,doi:10.1126/science.aaf2468}, and probing new physical forces and particles \cite{PhysRevA.108.052825}. In addition, a degenerate Fermi gas of ultracold deuterium atoms would enable a new paradigm in quantum simulation with the simplest Fermi liquid \cite{Zaghoo_PRL2019_metallic_deuterium}. Even without dissociation, the fermionic nature of CaD could allow it to reach lower temperatures and achieve better shielding from collisional losses in a conservative trap, creating a promising playground for quantum simulation experiments. The existence of an electron spin in the ground state of CaD provides an additional degree of freedom compared to bialkali molecules. The relative simplicity and extensive applications make CaD an interesting laser-cooling candidate. In this work, we demonstrate the production and one-dimensional (1D) laser cooling of CaD, which, to our knowledge, is the only fermionic molecule to have been directly cooled.

\section{Methods}

\begin{figure*}[ht!]
   \centering
   \includegraphics[scale=0.59]{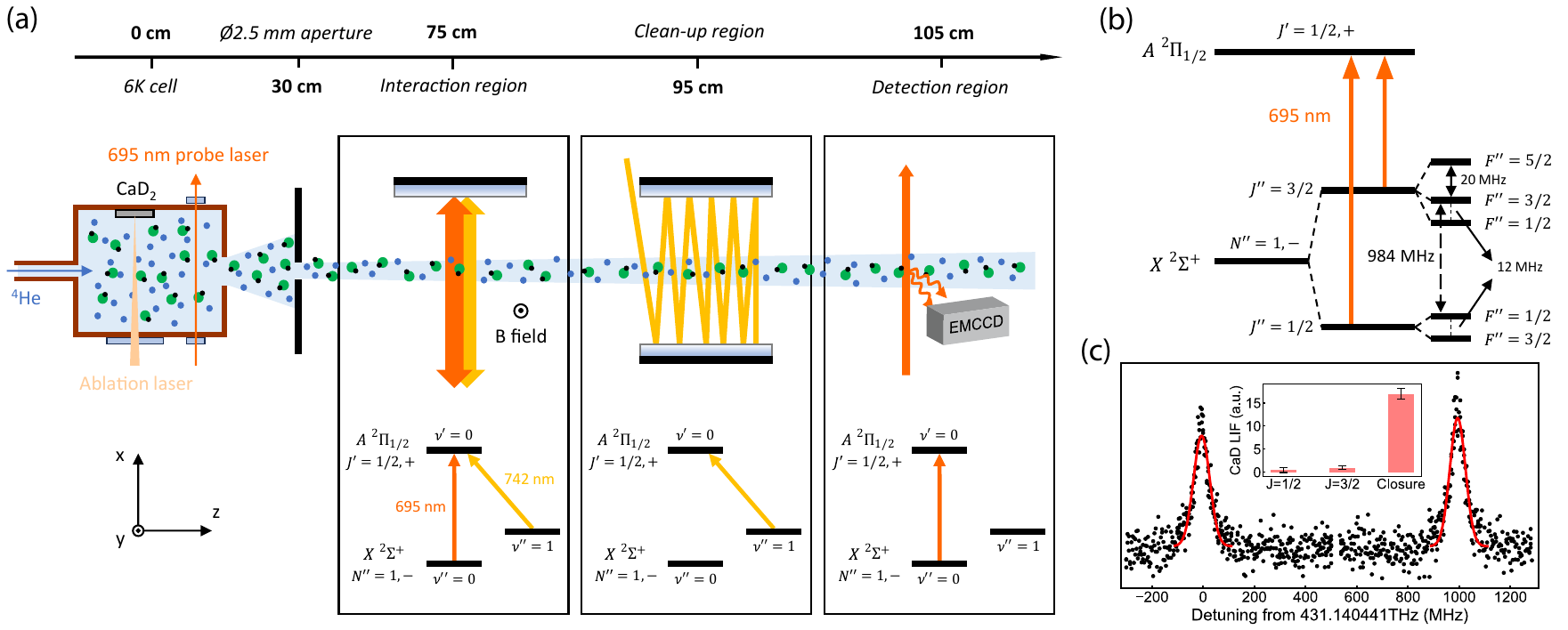}    
   \caption{(a) Schematic of the experiment viewed from above. CaD molecules are generated in a $\sim$6~K CBGB source. The molecular beam is collimated by a 2.5-mm diameter aperture and enters the interaction region where it is addressed by the main cooling laser, copropagating with the ($\nu=1$) repump laser. The laser beam is expanded to (10.8~mm$\times$5.4~mm) $1/e^2$ diameter and retroreflected to form a standing wave. The molecules then enter the clean-up region where only the ($\nu=1$) repump laser is applied in a multipass configuration and are finally detected in the ($\nu=0$) state. (b) Level structure of the main cycling transition $A^2\Pi_{1/2}(\nu'=0,\ J'=1/2,\ +)\leftarrow X^2\Sigma^+(\nu''=0,\ N''=1,\ -)$.  The hyperfine splittings are adapted from Ref. \cite{Frum_rotational_spectra_1993}. (c) In-beam spectroscopy of the transition. The inset shows signal enhancement with rotational closure.}
   \label{fig:schematic}
\end{figure*}

\subsection{Experimental setup}

Figure \ref{fig:schematic}(a) illustrates the experimental setup. The experiment starts with a cryogenic buffer-gas beam (CBGB) source. An Nd:YAG-pulsed laser at 1064~nm is used to ablate a solid rock sample of CaD$_2$, creating a hot cloud of CaD radicals. The molecules are buffer-gas cooled to $\sim$6~K using $^4$He and then ejected from the cell. This creates a CaD molecular-beam pulse with peak forward velocity of $\sim$250~m/s. The molecules are collimated with a 2.5-mm diameter aperture to 16(1)~mK transverse temperature (Appendix \ref{sec:transverse_temperature}) before entering the interaction region where we perform laser cooling. The molecules then enter the clean-up region and are detected through laser-induced fluorescence (LIF) with an electron-multiplying charge-coupled device (EMCCD) camera. The setup has been described in detail in the context of CaH experiments \cite{Vasquez-Carson_2022_CaH,Sun_PRR2023_Predissociation} with some modifications. Our ultrahigh-vacuum chamber is set up in a three-dimensional (3D) radio-frequency magneto-optical trap (RF-MOT) configuration, with the MOT chamber here serving as the interaction region. We apply magnetic field using the in-vacuum MOT coils in the Helmholtz configuration.

Although some deuterated compounds are commercially available, CaD$_2$ is not. We produce the molecules by following a procedure outlined in Ref. \cite{Bulanov2004synthesis} and detailed in Appendix \ref{sec:synthesis}. The flux of the CaD molecules in the $X^2\Sigma^+ (\nu''=0,\ N''=1,\ -)$ state out of the CBGB source is $\sim10^{11}$~molecules$/$steradian$/$pulse. Compared to CaH, we find an overall $\sim5\times$ higher yield under a full range of experimental parameters such as helium flow rate, ablation laser energy, and ablation laser-beam waist. This could be attributed to differences in collisional cross section with $^4$He and in crystalline chemistry between the species (Appendix \ref{sec:characterization}).

\subsection{Relevant level structure for laser cooling}

Figure \ref{fig:schematic}(b) shows the level structure of the main cycling transition $A^2\Pi_{1/2}(\nu'=0,\ J'=1/2,\ +)\leftarrow X^2\Sigma^+(\nu''=0,\ N''=1,\ -)$ used for laser cooling. We are able to locate the transition in Fig. \ref{fig:schematic}(c) within $\sim$200~MHz of available spectroscopic data \cite{Bernath2012spectrum,Alavi2018einsteincoeff}. The hyperfine splittings of the $J=1/2$ and $J=3/2$ states are adapted from Ref. \cite{Frum_rotational_spectra_1993}, which we are unable to resolve due to Doppler broadening of the beam. This is of particular advantage compared to CaH where the spacings are $\sim$50$-$100~MHz \cite{Vasquez-Carson_2022_CaH}. We show our ability to rotationally close the transition by adding the two spin-rotation components $J=1/2$ and $J=3/2$ in the inset of Fig. \ref{fig:schematic}(c). The LIF is normalized to that of $J=3/2$ only and we observe a $\sim15\times$ higher fluorescence when both components are present. This also confirms that the $A^2\Pi_{1/2}(\nu'=0,\ J'=1/2,\ +)$ excited-state hyperfine levels are unresolved.

A key requirement to directly laser cool a molecule is the ability to continuously scatter photons at a relatively fast rate ($\sim$10$^6$~s$^{-1}$). The vibrational branching ratios (VBRs) of the molecule should be diagonal so that a practical number of lasers can mitigate loss to higher vibrational states. The VBRs for CaD have been calculated \cite{Alavi2018einsteincoeff} and are shown in Table \ref{table:VBRs} of Appendix \ref{sec:VBRs}. Since the electronic potential energy surface does not depend on the mass of the nuclei, we do not expect large isotope shifts in the VBRs. However, the mass change leads to a significant shift of fundamental vibrational frequencies, from 37 THz in CaH to 27 THz in CaD. Overall, we assume the VBRs for CaD to be the same as our measured VBRs for CaH in order to calculate the number of scattered photons.

\begin{figure*}[ht!]
   \centering
   \includegraphics[scale=0.47]{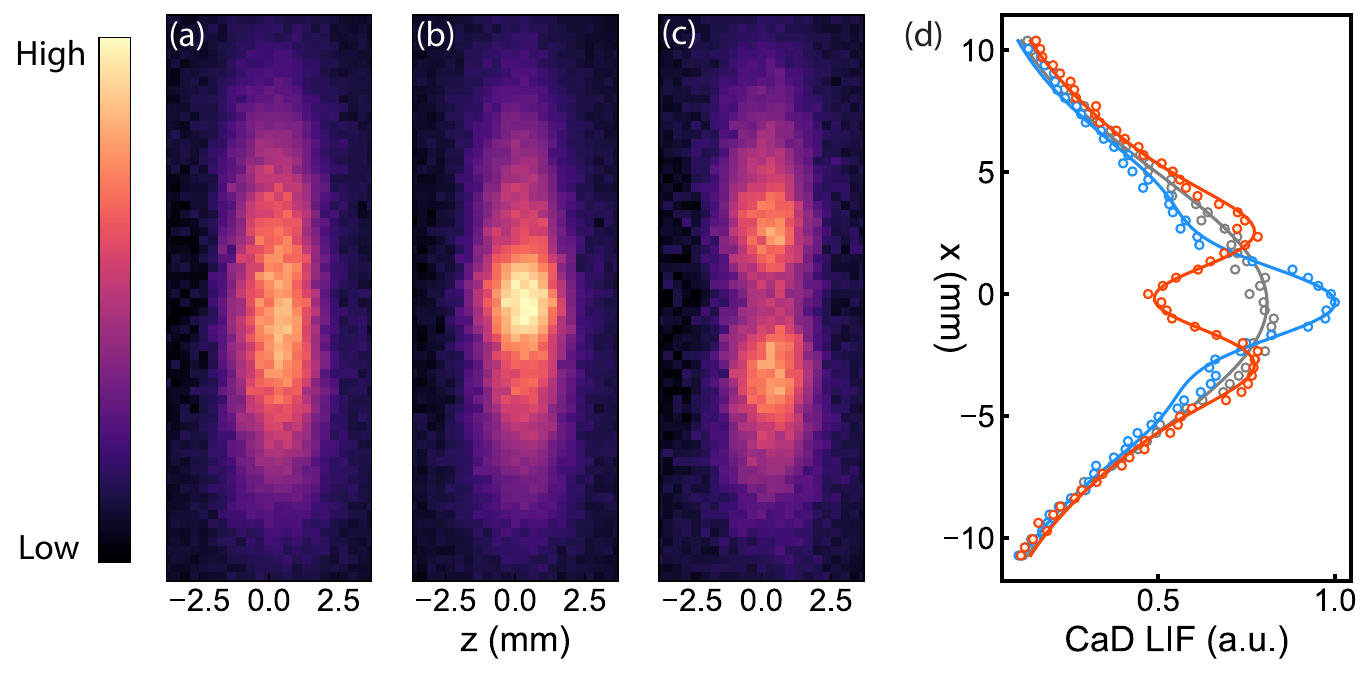}    
   \caption{Molecular-beam profile with Sisyphus cooling and heating. (a)$-$(c) Beam images under unperturbed, Sisyphus cooling at $+$40~MHz, and Sisyphus heating at $-$40~MHz configurations, respectively. (d) Integrated 1D profile. Sisyphus cooling narrows the width of the center of the molecular beam (blue), while Sisyphus heating expels molecules away from the center (red). Unperturbed beam is shown as gray. Solid lines show the fits to the beam profile.}
   \label{fig:beam_profile}
\end{figure*}

We perform laser cooling on the $A^2\Pi_{1/2}(\nu'=0,\ J'=1/2,\ +)\leftarrow X^2\Sigma^+(\nu''=0,\ N''=1,\ -)$ transition, together with a $(\nu=1)$ repump laser addressing the $A^2\Pi_{1/2}(\nu'=0,\ J'=1/2,\ +)\leftarrow X^2\Sigma^+(\nu''=1,\ N''=1,\ -)$ transition. This allows us to scatter $\sim1/(1-\text{VBR}_{\nu=0,\ 1})\approx400$~photons before populating higher vibrational states. We add frequency sidebands for the ground-state spin-rotation and hyperfine splittings using electro-optic modulators (EOMs) (see Appendix \ref{sec:laser_configuration} for details). The main cooling laser and the repump laser are combined with a polarizing beam splitter and coupled to a single-mode polarization-maintaining optical fiber. They are expanded to ($2w_1\times 2w_2$)$=$(10.8~mm$\times$5.4~mm) $1/e^2$ diameter and retroreflected in the interaction region to form a standing wave. The polarization of the main cooling laser ($\nu=0$) is set to 45$^{\circ}$ with respect to the applied magnetic field along the $y$ axis for efficient dark-state remixing. In the clean-up region, only the $(\nu=1)$ repump laser is applied in a multipass configuration. Finally we detect the molecules that remain in the $(\nu=0)$ state using the main cycling transition. The total power of the lasers shown in the schematic are 475~mW, 180~mW, 180~mW, and 45~mW, corresponding to the main cooling laser, $(\nu=1)$ repump laser in the interaction region, $(\nu=1)$ repump laser in the clean-up region, and the detection laser, respectively.

\section{Results}

\begin{figure*}[ht]
   \centering
   \includegraphics[scale=0.42]{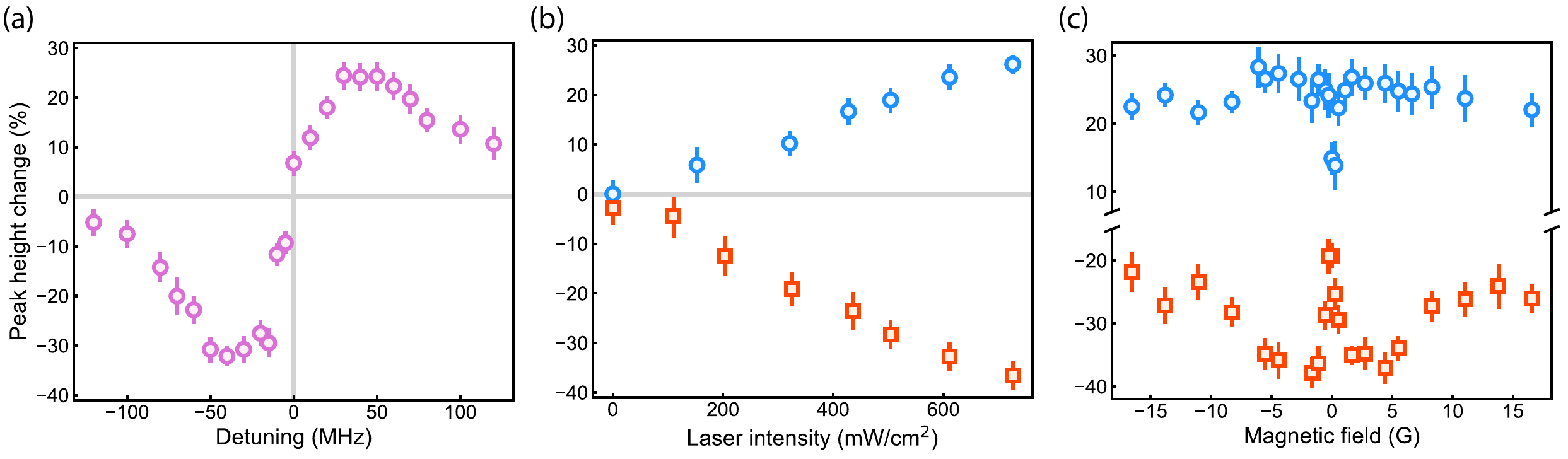}
   \caption{Parameter scans of the Sisyphus effect. (a) Peak height change as a function of the main laser detuning, taken with magnetic field $|\vec{B}|$ = 1.7~G and at 45$^\circ$ with respect to laser polarization. (b) Sisyphus cooling (blue circles) and heating (red squares) as a function of the main laser intensity, taken at $\pm$40~MHz detuning and the same magnetic field configuration as in (a). Saturation is not observed. The gray lines are guides to the eye for the unperturbed scenario (horizontal) and zero detuning [vertical in (a)]. (c) Sisyphus cooling (blue circles) and heating (red squares) as a function of applied magnetic field, taken at $\pm40$ MHz detuning and maximum laser intensity. A sharp decrease in the strength of Sisyphus effect is observed when $|\vec{B}|\approx0$. Error bars represent the 95$\%$ confidence intervals.}
   \label{fig:parameter_scan}
\end{figure*}

We demonstrate 1D cooling in a beam of CaD molecules using the magnetically assisted Sisyphus effect \cite{Emile_1993_sisyphus}. This technique has served as a barometer for the feasibility of laser cooling for molecules such as SrF \cite{Shuman_LaserCoolingDiatomic_2010},YbF \cite{Lim_CooledYbF_2018}, SrOH \cite{Kozyryev_SisyphusSrOH_2017}, YbOH \cite{Augenbraun_NJP2020_YbOH_Sisyphus}, and CaOCH$_3$ \cite{Mitra_CaOCH3Sisphus_2020}. Figures \ref{fig:beam_profile}(a)$-$\ref{fig:beam_profile}(c) show the molecular-beam profile detected on the EMCCD camera, with a magnetic field $|\vec{B}|$ = 1.7~G applied along the $y$ axis in the interaction region. The image in Fig. \ref{fig:beam_profile}(a) is taken when the main cooling laser is absent, representing the unperturbed molecular-beam profile in the $(\nu=0,\ 1)$ states. The fluorescence along the $x$ axis indicates the size of the molecular beam in the detection region, while the $z$-axis width is determined by the detection laser beam. In Fig. \ref{fig:beam_profile}(b) the main cooling laser is applied at $+$40 MHz detuning from resonance and we observe an accumulation of molecules near the center of the beam due to Sisyphus cooling.  In Fig. \ref{fig:beam_profile}(c) the laser is detuned by $-$40~MHz and we observe an expulsion of molecules from the center due to Sisyphus heating. We integrate the images along the $z$ axis and obtain the 1D profile shown in Fig. \ref{fig:beam_profile}(d). We note that because of the limited interaction time of $\sim40$~$\mu s$, we only deplete the population to 93(3)\% even when the main cooling laser is set on resonance. This corresponds to 30(13)~photons scattered at a rate of $0.7(3)\times 10^6$~s$^{-1}$. This number further decreases in Sisyphus cooling and heating configurations due to the laser detuning. With the limited number of scattered photons, the integrated LIF signals under Sisyphus cooling and heating configurations are essentially unchanged from an unperturbed beam, as in Fig. \ref{fig:beam_profile}(d). We note that the estimated scattering rate is roughly half of the maximum scattering rate due to the fact that both $(\nu=0)$ and $(\nu=1)$ lasers address the same excited state $A^2\Pi_{1/2}(\nu'=0,\ J'=1/2,\ +)$.

\subsection{Fitting protocol}

We fit the 1D profile with a phenomenological model capable of describing both Sisyphus cooling and heating. The model combines a Gaussian and a second derivative of a Gaussian, the latter serving as a modulation of the amplitude. The function is given by
\begin{equation}
f(x)=A\times g_{x_0,\sigma_0}(x)\times(1-\Delta h\times g''_{x_0,\sigma_S}(x)),
\label{eq:fit}
\end{equation}
where $g_{x_0,\sigma_0}(x)=\exp(-(x-x_0)^2/(2\sigma_0^2))$ and $g''_{x_0,\sigma_S}(x)=((x-x_0)^2/\sigma_S^2-1)\times \exp(-(x-x_0)^2/(2\sigma_S^2))$. The parameters $A$, $x_0$, and $\sigma_0$ represent the amplitude, center position, and width of an unperturbed beam, while $\Delta h$ and $\sigma_S$ represent the fractional peak height change and the width of the region where Sisyphus cooling or heating is effective. The shape of the second derivative of a Gaussian qualitatively reproduces how Sisyphus cooling or heating affects the central part of the molecular beam. The fits are shown as solid lines in Fig. \ref{fig:beam_profile}(d).

The characteristic feature of the Sisyphus effect arises from the finite capture velocity $v_c$ of the Sisyphus force. In the Sisyphus cooling configuration, only molecules with $v<v_c$ can experience the Sisyphus force and get cooled. The Sisyphus cooled molecules reach a transverse temperature of 2.7(2)~mK. In the Sisyphus heating configuration, on the contrary, slower molecules are heated and pushed away from the center. From the position of the two peaks in the Sisyphus heating profile, we determine a capture velocity of $\sim$1~m$/$s.

\subsection{Characterization of the Sisyphus effect}

We characterize the Sisyphus effect for the following experimental parameters: laser detuning, intensity, and magnetic field. The Sisyphus effect as a function of laser detuning is shown in Fig. \ref{fig:parameter_scan}(a), taken at $|\vec{B}|$=1.7~G. Here we use peak height change $\Delta h$ in Eq. (\ref{eq:fit}) as a proxy for the strength, its sign representing cooling (positive) or heating (negative). The Sisyphus effect is antisymmetric around zero detuning, and the optimal detuning for cooling and heating are found to be at $\sim\pm$40~MHz, respectively, which corresponds to $\sim\pm$10$\Gamma$. Here $\Gamma \approx $$(2\pi)\times$3.8~MHz represents the natural linewidth of the main cycling transition and is estimated from Ref. \cite{Alavi2018einsteincoeff}. The relatively large optimal detunings primarily result from the small hyperfine splittings, since at small detunings all the sidebands introduced to address the hyperfine structure of the ground state compete with each other. Figure \ref{fig:parameter_scan}(b) shows the laser-intensity dependence of Sisyphus cooling (blue circles) and heating (red squares). Laser intensity here is defined as the mean intensity of the Gaussian beam in a $1/e^2$-diameter area $(P_{\text{tot}}/\pi w_1w_2)$, with only the effective laser sidebands accounted (Appendix \ref{sec:laser_configuration}). We are not able to saturate the Sisyphus effect primarily due to the limited interaction time, and the scattering rate increases linearly with laser intensity in this regime.

The Sisyphus mechanism in a Type-II system with more ground states than excited states relies on dark-state remixing. In the cooling configuration, for example, molecules lose kinetic energy as they climb to the intensity maxima of the standing wave produced by the retroreflected laser beam. At the maxima they are optically pumped into dark states at a lower energy \cite{Emile_1993_sisyphus}. Dark-state remixing allows the molecules to return to a bright state and repeat the cycle, thus reducing the beam temperature. This is achieved through Larmor precession in the hyperfine sublevels in the presence of an external magnetic field. Figure \ref{fig:parameter_scan}(c) shows the magnetic-field response of the Sisyphus effect, taken at maximum laser intensity and $\pm$40~MHz detuning for cooling (blue circles) and heating (red squares). We observe a sharp decrease of strength in both cooling and heating at $|\vec{B}|\approx0$, when dark-state remixing is the least efficient. At higher magnetic fields, Sisyphus effect is again suppressed due to large Zeeman shifts. We also note that the Sisyphus effect cannot be completely eliminated by tuning the magnetic field in the standing-wave configuration for two reasons. First, the magnetic field cannot be perfectly canceled throughout the entire interaction region due to the presence of the Earth's magnetic field. Second, polarization gradient cooling could occur due to imperfect linear polarizations.

\section{Conclusion}

In conclusion, we have demonstrated laser cooling of a fermionic molecule, CaD. The complexity introduced by the larger nuclear spin $I_{\text{D}}=1$ is comparable to that of other bosonic molecules that have been laser cooled and trapped. This is a key first step towards a MOT of a fermionic molecule, opening the door to a variety of quantum science applications such as the realization of the dipolar Fermi-Hubbard model \cite{vanLoon_2015PRB_dipolar_FermiHubbard} and topological superfluids \cite{Cooper_2009_topological_superfluid}. With a fermionic molecule added to the list of directly laser-coolable molecular species, qudit platforms with high-fidelity readout become possible \cite{Sawant_UltracoldMolQubit_2020}. Due to a small isotope shift between CaD and CaH, we expect the deuterated isotopologue to suffer from similar rates of predissociation for the $A$ and $B$ excited states as CaH \cite{Sun_PRR2023_Predissociation}. However, the same predissociative nature can be leveraged to engineer dissociation pathways for the generation of ultracold and trapped atomic deuterium \cite{Lane_UltrcoldHydrogen_2015}. This will enable a novel precision-measurement platform with trapped hydrogen and deuterium \cite{PhysRevA.108.052825}. Finally, the efficient hydrogenation and deuteration method described in this work can be extended to produce chiral molecular isomers such as the chiral methyl group \cite{Floss_1993_chiral_methyl} in order to study parity-violation effects predicted to be the source of biomolecular homochirality~\cite{Quack_Chiral_Matter}.

\section*{Acknowledgments}

We would like to thank J. S. Owen, J. R. Caram, M. J. Nava, and D. Tencio for useful discussions about the feasibility of the chemical reaction. We thank J. Geisenhoff for performing the PXRD measurement on the sample. We also thank L. Cheng for discussions about isotope shift effects on branching ratios. We are grateful to C. Hallas and J. MacArthur for help with the 16-channel wavemeter setup.  This work was supported by the ONR Grant No. N00014-21-1-2644, AFOSR MURI Grant No. FA9550-21-1-0069, AFOSR DURIP Grant No. FA9550-23-1-0149, and we acknowledge generous support by the Brown Science Foundation.

\appendix
\section{SYNTHESIS OF CaD$_2$}
\label{sec:synthesis}

\begin{figure}[ht!]
   \centering
   \includegraphics[scale=1]{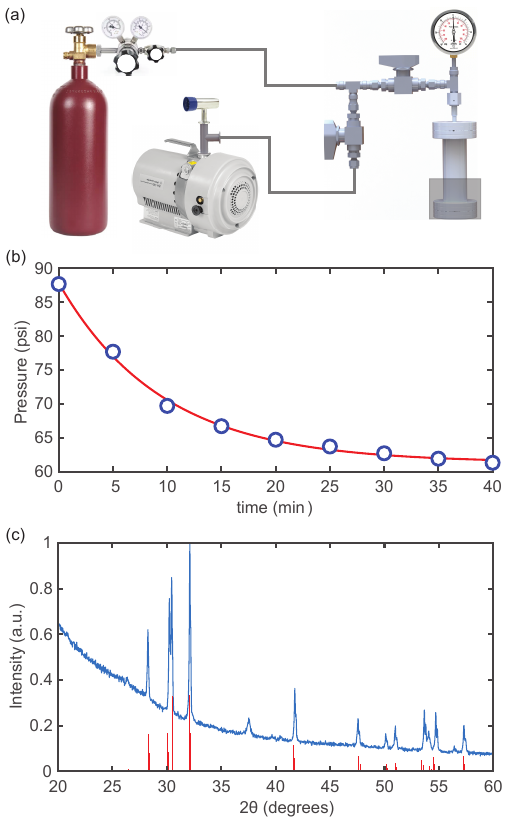}    
   \caption{Synthesis of CaD$_2$. (a) Illustration of the home-built setup used for the synthesis. The main reactor is a standard CF 2.75", 5" long 304 stainless steel nipple. The base is wrapped with heating tape connected to a variac via a temperature controller (Extech 48VFL). A standard J-type thermocouple attached to the base is used for temperature stabilization. A high-pressure gauge (MG1-100-A-9V-R) is placed close to the top of the reactor. All connections henceforth are VCR type. The reactor connects to a scroll pump and a Convectron gauge (MKS 275) via a tee. The other end of the tee connects to a D$_2$ cylinder using a long flexible hose. The regulator on the cylinder allows for control of the charging pressure, while the valves shown are used to direct the flow. (b) Pressure measured on the high-pressure gauge as a function of time at 450 $^{\circ}$C. The reactor contains 3 g of Ca pieces. From an exponential fit, we obtain a reaction rate of $k =$ 0.11~min$^{-1}$. (c) Powder x-ray diffraction study of the resulting sample. Red lines are the known CaD$_2$ peaks. Almost all measured peaks match with the known CaD$_2$ values except for the peak at $2\theta=37^{\circ}$. This peak could be attributed to CaO or Ca(OH)$_2$, possibly resulting from short contact with air. The results suggest very high conversion efficiency from Ca to CaD$_2$.}
   \label{fig:synthesis}
\end{figure}

We produce CaD$_2$ samples by following a procedure outlined in Ref. \cite{Bulanov2004synthesis} for the production of CaH$_2$. We build a high pressure, high temperature reactor using readily available components [Fig. \ref{fig:synthesis}(a)]. The body of the reactor is composed of a standard CF 2.75" stainless steel nipple. The bottom of the nipple is blanked off using a nickel gasket and constitutes the main reaction area wrapped with an electrical heating coil and a thermocouple connected to a temperature controller. The top is a CF to 1/4" VCR adapter and the subsequent connections are all made with standard stainless steel VCR connectors. Such connections are rated for pressures $>5,000$ psi and temperatures $>500$ $^\circ$C. However, CF flanges are only rated for pressures below atmospheric pressure and temperatures up to 450 $^\circ$C.

Once assembled, we stress test the reactor using inert helium gas. We confirm that the system is leak-tight first at atmospheric pressure and 450~$^\circ$C, then up to 80 psi at room temperature, and finally up to 80 psi at 450 $^\circ$C. With no leaks appearing at any point during these tests, we deem our reactor safe for operation with deuterium gas. We use calcium pieces of >99\% purity from Millipore Sigma (327387-25G) and 99.8\% pure deuterium gas from Cambridge Isotope Labs (DLM-408-100). 

We charge our reactor with 3 g of Ca metal pieces measuring a few millimeters in size. Powderizing the Ca target is expected to improve reaction yield due to larger surface area but this is avoided since it would require pressing to form a target. The reactor is charged and sealed in an inert nitrogen environment to avoid contamination. Then we gradually heat the sample under vacuum to 450 $^\circ$C over 2 hours. Care is taken to avoid large thermal gradients in the system that could lead to a formation of leaks. Since Ca melts at 840 $^\circ$C, this temperature only enhances the rate of deuterating the Ca pieces. We then charge the hot reactor instantly with D$_2$ gas to 88 psi. With the D$_2$ reservoir valved off, the pressure in the reactor starts decreasing immediately, signaling that the reaction is underway. The pressure decreases to 62 psi after 40 min and the rate of change stagnates. This implies that the first phase of the reaction, surface deuteration, is complete [Fig. \ref{fig:synthesis}(b)]. The measured reaction rate is $k$ = 0.11~min$^{-1}$. We leave the reactor charged and hot overnight, and after 9 hours the pressure decreases to $\sim$50 psi. During this time, D$_2$ molecules diffuse through the Ca surface and penetrate into the bulk.

Once the system has cooled down to room temperature, we open the reactor in a nitrogen environment. Since the temperature is not high enough to melt calcium, the pieces do not change their shape or size. However, they turn from a dark silvery gray to a light powdery gray. We perform a powder x-ray diffraction (PXRD) analysis of the sample and find that it predominantly consists of CaD$_2$ [Fig. \ref{fig:synthesis}(c)]. This confirms our assumption that D$_2$ molecules can easily diffuse through the metal surface and cause deuteration in the bulk of the Ca pieces with high efficiency.

\section{CHARACTERIZATION OF MOLECULAR YIELD}
\label{sec:characterization}

We employ a CBGB source described in detail in Ref. \cite{Vasquez-Carson_2022_CaH}. The steady-state in-cell density of the He buffer gas is given by Eq. (\ref{bgasdensity}), where $f_\text{He}$ is the flow rate of the buffer gas and $v_\text{He}$ is its average thermal velocity at 6~K \cite{Hutzler_CR2012_BufferGasBeams}:
\begin{equation}
\text{n}_{\text{He}} = \frac{4f_{\text{He}}}{A_{\text{aperture}}v_{\text{He}}}.
\label{bgasdensity}
\end{equation} 
For the measurements taken in this work a buffer-gas flow rate of $\sim$20 standard cubic centimeters per minute (sccm) is used, giving a steady-state buffer-gas density of 1.4$\times 10^{16}\text{ cm}^{-3}$, similar to other buffer-gas experiments. Although the higher He flow rate leads to higher beam velocities, we obtain a higher molecular yield and signal-to-noise ratio (SNR). Interestingly, we measure a $\sim 5\times$ higher CaD yield both inside the buffer-gas cell and also in the extracted beam compared to CaH measured under identical conditions.

This difference in production of the two isotopologues can give insight into the dynamics of our CBGB source, allowing for further improvements in molecule production and molecular-beam flux. Differences in thermalization resulting from different collisional cross sections is one possible explanation for our observations. As the hot molecules diffuse away from the target, they collide with the cold buffer gas. The rate of these collisions affects the time it takes the molecules to diffuse through the cell. Assuming the molecular species instantly freezes (or is ``lost'') once it reaches the cell walls, and the molecules are much lower density than the buffer gas, the decay in absorption is a good representation of the time it takes for the molecules to diffuse out. This characteristic time $\tau$ can be found by fitting a simple exponential to the tails of absorption traces. Then, using the method employed in Ref.~\cite{Hutzler_CR2012_BufferGasBeams}, the collisional cross section can be calculated via the expression
\begin{equation}
    \sigma_{\text{He-m}} = \tau \frac{9\pi v_{\text{He}}}{16 A_{\text{cell}}\text{n}_{\text{He}}},
\end{equation}
where $A_{\text{cell}}$ is the cross-sectional area of the cell, $\text{n}_{\text{He}}$ is the steady-state density from Eq. (\ref{bgasdensity}), and $v_{\text{He}}$ is the thermal velocity of the buffer gas.  We find the cross sections to be $\sigma_{\text{He-CaH}} = 2.0(8)\times 10^{-14}~\text{cm}^2$ and $\sigma_{\text{He-CaD}} = 3.0(3)\times 10^{-14}~\text{cm}^2$. These calculated cross sections are consistent with previous work \cite{weinstein2002magnetic}. However, given the large errors of these values and their relative closeness, we cannot definitively attribute the difference in the production between the two species to He thermalization. We also cannot rule out other factors such as crystalline chemistry of the substrates.

\section{VIBRATIONAL BRANCHING RATIOS}
\label{sec:VBRs}

The VBRs for CaD calculated from Ref. \cite{Alavi2018einsteincoeff} are shown in Table \ref{table:VBRs}. We also show the calculated CaH VBRs and find a small isotope shift between CaH and CaD. We make the assumption that the VBRs for CaD are approximately the same as those for CaH.

\begin{table}[]
\centering
\begin{tabular}{|c|c|c|c|c|}
\hline
Transition & \begin{tabular}[c]{@{}c@{}}Vibrational \\ Quanta \\ $(\nu'')$\end{tabular} & \begin{tabular}[c]{@{}c@{}}CaH VBR \\ Measured \\ $(q_{0\nu''})$\end{tabular} & \begin{tabular}[c]{@{}c@{}}CaH VBR \\ Calculated \\ $(q_{0\nu''})$\end{tabular} & \begin{tabular}[c]{@{}c@{}}CaD VBR \\ Calculated \\ $(q_{0\nu''})$\end{tabular} \\ \hline
\multirow{4}{*}{$A\rightarrow X$} & 0 & 0.9680(29) & 0.9820 & 0.9758 \\ \cline{2-5} 
 & 1 & 0.0296(24) & 0.0175 & 0.0235 \\ \cline{2-5} 
 & 2 & 2.4(1.8)$\times 10^{-3}$ & 4.61$\times 10^{-4}$ & 7.18$\times 10^{-4}$ \\ \cline{2-5} 
 & 3 & - & 2.0$\times 10^{-5}$ & 3.2$\times 10^{-5}$ \\ \hline
\multirow{4}{*}{$B\rightarrow X$} & 0 & 0.9853(11) & 0.9790 & 0.9688 \\ \cline{2-5} 
 & 1 & 0.0135(11) & 0.0202 & 0.0295 \\ \cline{2-5} 
 & 2 & 1.2(0.2)$\times 10^{-3}$ & 7.5$\times 10^{-4}$ & 1.5$\times 10^{-3}$ \\ \cline{2-5} 
 & 3 & - & 6.7$\times 10^{-5}$ & 1.5$\times 10^{-4}$ \\ \hline
\end{tabular}
\caption{Measured CaH VBRs \cite{Vasquez-Carson_2022_CaH} and calculated CaH and CaD VBRs \cite{Alavi2018einsteincoeff}. The calculated VBRs for CaH are in good agreement with our own calculations \cite{Sun_PRR2023_Predissociation}.}
\label{table:VBRs}
\end{table}

\section{LASER CONFIGURATION}
\label{sec:laser_configuration}

The frequencies of the lasers used in this work are presented in Table \ref{table:laser_frequencies}. The spectroscopy of the $A^2\Pi_{1/2}(\nu'=0,\ J'=1/2,\ +)\leftarrow X^2\Sigma^+(\nu''=0,\ N''=1,\ -)$ main cycling transition is done in a quasirotationally-closed configuration. We scan the frequency of one spin-rotation component while holding the other on resonance. This provides higher SNR because of the enhancement in LIF due to optical cycling. With the hyperfine levels unresolved [Fig. \ref{fig:schematic}(c)], we fit the spectra to Gaussians with fixed spacing and relative amplitude, according to the hyperfine splittings obtained from Ref. \cite{Frum_rotational_spectra_1993} and state degeneracies.  However, there are no previously measured hyperfine splittings for the $X^2\Sigma^+(\nu''=1,\ N''=1,\ -)$ state. We instead fit the $(\nu=1)$ repump transition with a single Gaussian for each spin-rotation component. All the frequencies are measured using a HighFinesse WS7-60 wavemeter, which is calibrated using the $^1S_0\rightarrow ^3P_1$ transition in Sr.

\begin{table}[ht]
\centering
\small\addtolength{\tabcolsep}{-1pt}
\begin{tabular}{|c|c|c|c|c|c|c|c|c|c|}
\hline
Ground             & $\nu''$                & $N''$                & $J''$                  & $F''$           & Excited            & $\nu'$                 & $N'$                 & $J'$                   & Frequency (THz) \\ \hline
\multirow{5}{*}{$X$} & \multirow{5}{*}{0} & \multirow{5}{*}{1} & \multirow{3}{*}{3/2} & 5/2           & \multirow{5}{*}{$A$} & \multirow{5}{*}{0} & \multirow{5}{*}{-} & \multirow{5}{*}{1/2} & 431.140421      \\ \cline{5-5} \cline{10-10} 
                   &                    &                    &                      & 3/2           &                    &                    &                    &                      & 431.140441      \\ \cline{5-5} \cline{10-10} 
                   &                    &                    &                      & 1/2           &                    &                    &                    &                      & 431.140453      \\ \cline{4-5} \cline{10-10} 
                   &                    &                    & \multirow{2}{*}{1/2} & 3/2           &                    &                    &                    &                      & 431.141437      \\ \cline{5-5} \cline{10-10} 
                   &                    &                    &                      & 1/2           &                    &                    &                    &                      & 431.141425      \\ \hline
\multirow{2}{*}{$X$} & \multirow{2}{*}{1} & \multirow{2}{*}{1} & 3/2                  & 5/2, 3/2, 1/2 & \multirow{2}{*}{$A$} & \multirow{2}{*}{0} & \multirow{2}{*}{-} & \multirow{2}{*}{1/2} & 403.852444      \\ \cline{4-5} \cline{10-10} 
                   &                    &                    & 1/2                  & 3/2, 1/2      &                    &                    &                    &                      & 403.853423      \\ \hline
\end{tabular}
\caption{CaD transition frequencies used in this work. The uncertainties are $\sim$10 MHz statistical and $\sim$60 MHz systematic from the wavemeter.}
\label{table:laser_frequencies}
\end{table}

In order to generate the sidebands for the ground-state spin-rotation and hyperfine splittings of the main cycling transition, we use two EOMs operating at 980~MHz and 11~MHz in series.  Specifically, we use the $-$1st and 0th order of the 980~MHz EOM for the $J=3/2$ and $J=1/2$ spin-rotation components, respectively, and the 0th, $\pm$1st, $\pm$2nd orders of the 11~MHz EOM for the hyperfine levels. We note that the $+$1st order of the 980~MHz EOM is unused in the cooling experiment because it is far-off resonant, resulting in a $\sim$30$\%$ decrease in the effective laser intensity. For the $(\nu=1)$ repump, we use a 978~MHz EOM and a high-Q 4.185~MHz EOM. The 4.185~MHz EOM generates an array of sidebands ($\pm5$ orders) to cover the hyperfine levels in the $(\nu=1)$ manifold.

\section{BEAM TEMPERATURE ESTIMATION}
\label{sec:transverse_temperature}

We estimate the transverse temperature of the molecular beam using Monte Carlo beam propagation. We generate a uniform spatial distribution of $10^5$ particles at the location of the aperture with a Boltzmann-distributed forward velocity of 250~m/s with 40~m/s as standard deviation and a transverse velocity $v_\perp$. We vary $v_\perp$ and calculate the molecular-beam width at the position of the detector. Matching this value to our measured unperturbed beam width ($\sigma\approx5.4$~mm) gives us an estimate of $T_\perp\approx$~16(1)~mK. Next, we perform the same computation but now compare the calculated beam width to the measured width of the central peak of the Sisyphus cooled profile [Fig. \ref{fig:beam_profile}(d)], $\sigma\approx$ 2.3~mm. This provides a rough estimate of the Sisyphus-cooled beam temperature of 2.7(2)~mK. We note that this temperature estimate is approximate and only serves as an indicator of cooling efficiency.

\vspace{23pt}
\section{DOPPLER COOLING}
\label{sec:doppler}

\begin{figure}[htpb]
   \centering
   \includegraphics[scale=0.5]{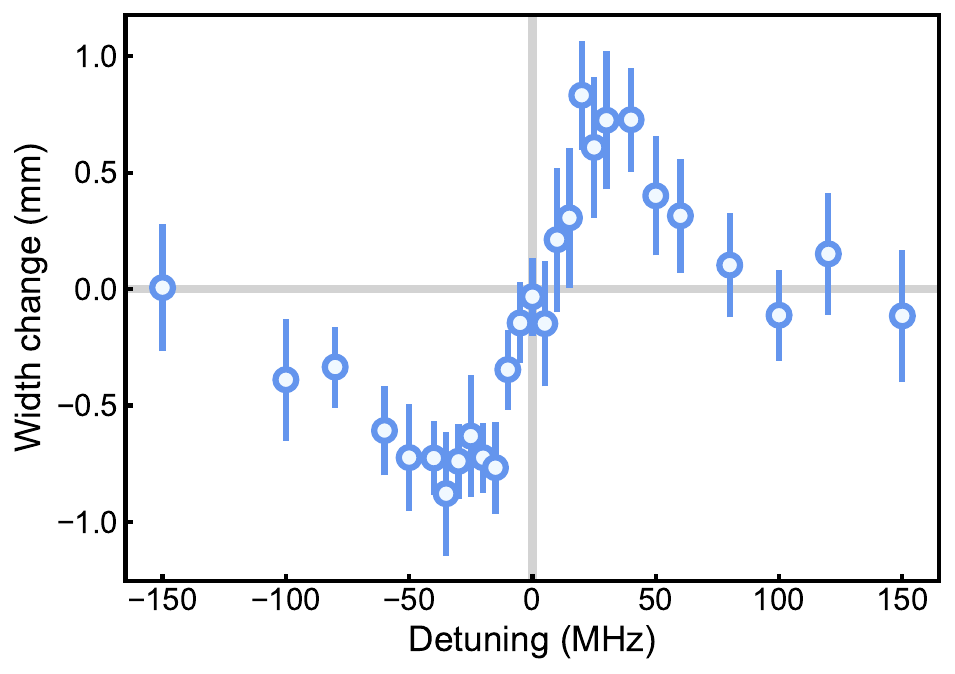}
   \caption{Doppler cooling and heating as a function of laser detuning, A change in the molecular-beam width demonstrates Doppler cooling (negative) and heating (positive). Error bars represent the 95$\%$ confidence intervals.}
   \label{fig:doppler}
\end{figure}

In addition to sub-Doppler Sisyphus cooling, we also perform 1D Doppler cooling of the CaD molecular beam. The measured Doppler cooling and heating curve as a function of laser detuning is shown in Fig. \ref{fig:doppler}. Because of the large capture velocity of the Doppler force, cooling and heating manifest as a change in Gaussian width ($\Delta\sigma$) of the molecular beam. We can switch from a Sisyphus to a Doppler configuration by multipassing the interaction laser and ensuring little overlap between neighboring passes.

%

\end{document}